\documentclass[%
 reprint,
 amsmath,amssymb,
 aps,
 prl,
 floatfix,
]{revtex4-2}
\usepackage{graphicx}
\usepackage{dcolumn}
\usepackage{bm}
\usepackage[colorlinks=true, linkcolor=blue, citecolor=blue, urlcolor=blue]{hyperref}
\usepackage{booktabs}

\begin{document}

\title{Data-driven structural diagnostics and autonomous alignment of complex optical systems}

\author{Peng Zhu}
\affiliation{Advanced Materials, Hong Kong University of Science and Technology (Guangzhou), Guangzhou 511453, China}
\author{Chen Chen}
\affiliation{Advanced Materials, Hong Kong University of Science and Technology (Guangzhou), Guangzhou 511453, China}
\author{Shuchang Ma}
\affiliation{Advanced Materials, Hong Kong University of Science and Technology (Guangzhou), Guangzhou 511453, China}
\author{Dezhou Deng}
\affiliation{Advanced Materials, Hong Kong University of Science and Technology (Guangzhou), Guangzhou 511453, China}

\author{J. F. Chen}
\email{chenjf@sustech.edu.cn}
\affiliation{State Key Laboratory of Quantum Functional Materials, Department of Physics, Southern University of Science and Technology, Shenzhen 518055, China}

\author{Peng Chen}
\email{pengchen@hkust-gz.edu.cn}
\affiliation{Advanced Materials, Hong Kong University of Science and Technology (Guangzhou), Guangzhou 511453, China}
\affiliation{Quantum Science Center of Guangdong-Hong Kong-Macau Greater Bay Area, Shenzhen 518045, China}

\date{\today}

\begin{abstract}
The quest for autonomous alignment and maintenance of complex free-space optical systems is increasingly urgent for large-scale neutral-atom quantum processors. Leveraging a high-finesse optical cavity as a sensitive probe, we introduce a data-driven framework for structural diagnostics and closed-loop control that achieves cold-start convergence within tens of seconds. These results establish a rapid, scalable diagnostic-control methodology for autonomous alignment and continuous maintenance of complex free-space optical architectures.
\end{abstract}

\maketitle
Driven by breakthroughs in single-atom addressing, reconfigurable atom-by-atom assembly, and efficient quantum error correction strategies, neutral-atom quantum computing has emerged as a promising platform for scalable quantum processors~\cite{bluvstein2024logical,Endres2023nature,Saffman2023PRX,kaufman2021quantum,henriet2020quantum,bluvstein2026fault}. A key advantage of these platforms lies in their connectivity~\cite{Lukin2001PhysRevLett,Saffman2010RevModPhys,Bluvstein2022Nature} and scalability~\cite{sheng2022defect,manetsch2025tweezer}, enabled by precise free-space optical architectures. Such systems manipulate hundreds or even thousands of independently addressed atomic qubits through sophisticated optical arrays~\cite{chiu2025continuous,manetsch2025tweezer,Pan2025PRL_array} (e.g., spatial light modulators and acousto-optic deflectors), making their operation highly dependent on the spatiotemporal stability of complex optical configurations~\cite{savard1997laser}. However, such complex free-space optical systems are inherently vulnerable to the undetected misalignment of even a single component. Debugging is notoriously challenging, as a localized fault can compromise the entire beam path, necessitating time-intensive diagnostic procedures to isolate the source \cite{saha2025automating}. This challenge is further compounded by persistent environmental perturbations (e.g., mechanical vibrations and thermal drifts~\cite{bluvstein2024logical,kaufman2021quantum}) and long-term beam-pointing instabilities, which continuously exacerbate control errors~\cite{mishra2026data,choi2026framework}. 

While automated alignment remains feasible for low-DOF systems using conventional algorithms, traditional analytical models (e.g., paraxial ray transfer matrices) struggle to accurately capture realistic misalignment configurations.  In ``cold-start'' scenarios defined by gross optical misalignment, gradient-based optimizers (e.g., hill-climbing) converge poorly due to flat response regions and local optima~\cite{saha2025automating,acernese2015advanced,morris2024general}. Gradient-free alternatives, such as reinforcement learning (RL)~\cite{acernese2015advanced} and stochastic parallel gradient descent (SPGD)~\cite{hu2020adaptive}, similarly suffer from prohibitive computational overhead and sluggish convergence. This creates a critical bottleneck for neutral-atom platforms that require alignment across high-dimensional spaces, necessitating intensive manual intervention, particularly during the initial setup phase. 

This problem can be addressed by developing a practical diagnostic framework that extracts a quantitative map of the intrinsic coupling structure governing a complex optical system from purely data-driven observations. Such a diagnostic capability would fundamentally transform the maintenance and troubleshooting of multi-DOF hardware, shifting the paradigm from blind, experience-dependent trial-and-error to targeted intervention guided by mathematically identified sensitivity hierarchies.

Optical resonant cavities, ubiquitous in spatial-mode mapping, serve as an ideal diagnostic tool for this task~\cite{Cavity2024Phys_Rev_A}. As highly sensitive interferometric systems, they respond sharply to small perturbations in optical alignment. Even a slight angular misalignment of a mirror or a sub-wavelength axial shift of a mode-matching element can drastically alter the intracavity field, rapidly degrading mode purity and exciting higher-order transverse mode~\cite{Cavity2024Phys_Rev_Lett}. Moreover, the cavity transmission image serves directly as the diagnostic signal, condensing the cumulative misalignment of the entire upstream free-space optical network (e.g., beam-delivery systems in neutral-atom arrays) into a single spatially resolved observable, which  intrinsically encodes angular drifts, thermal lensing, and cross-parameter coupling. Consequently, the diagnostic signal exhibits a steep, multi-peaked response surface with strong parameter correlations, rendering conventional iterative alignment strategies highly susceptible to convergence failure.

In this work, we present a unified framework that integrates end-to-end deep learning
inversion with partial least squares (PLS) analysis to extract the intrinsic coupling
structure of the control space. The framework is validated experimentally on a 5-DOF
optical resonant cavity~\cite{wang2025automatic,drever1983laser}. Applying PLS to the inverse-mapped states
reveals that the nominally five-dimensional control space is concentrated in a
three-dimensional subspace, and this subspace accounts for 80.3\% of the explained
variance. The dominant PLS1 direction loads most strongly on the lens axial position and the downstream mirror angles. This sensitivity hierarchy guides a hybrid navigation protocol in which a neural-network global navigator first traverses the cold-start regime in an average of 2.25 iterations, and a diagnostics-informed linear scan then refines the alignment. Across 40 randomized trials, the protocol converges in every case, raising the mean mode purity to above 60\% in the global phase and stabilizing it above 90\% after the local scan within tens of seconds. An auxiliary reflected-beam feedback loop also continuously compensates slow environmental drifts, so the framework extends from initial alignment to ongoing maintenance. Extension to 9-DOF confirms robustness to structural redundancy, as the validation loss remains comparable to that of the closed-loop-validated 5-DOF case. Architecture-aware analysis further indicates that the required training budget grows approximately linearly with dimensionality, circumventing the exponential complexity of grid-search methods. By leveraging the optical resonant cavity as a highly sensitive
diagnostic probe for the upstream free-space optical network, this work thereby establishes a
generalizable framework for autonomous alignment and continuous maintenance of complex free-space optical architectures, addressing both the cold-start bottleneck and long-term drift.

 \begin{figure*}[htbp]
    \centering
    \includegraphics[width=\linewidth]{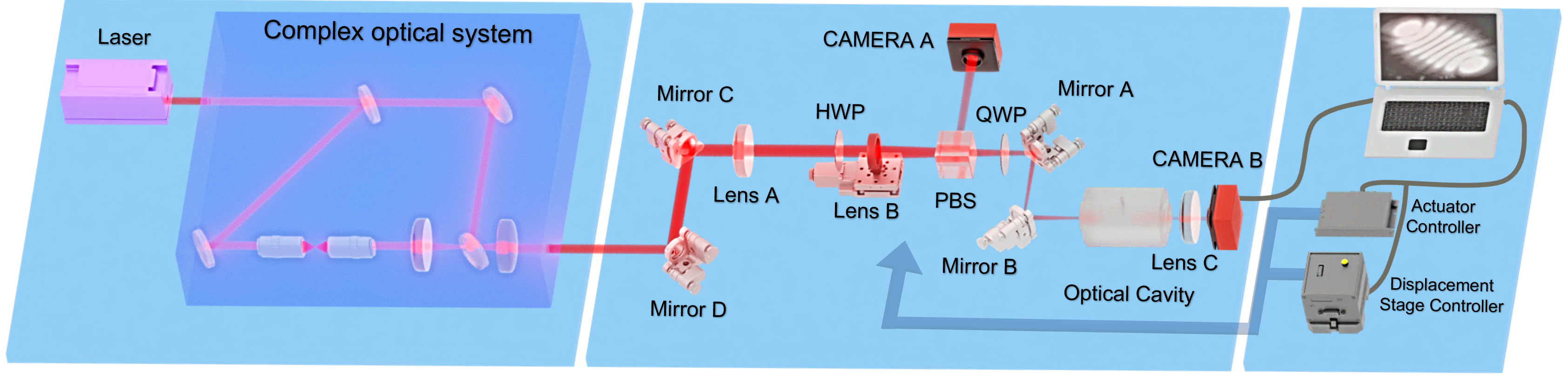}
\caption{$\mathbf{\vert}$\textbf{Schematic of the end-to-end automated alignment system.} The system (the middle and right panels) consists of three components: a free-space optical layout, a software-based data acquisition and processing unit, and an electronic control module. The cavity's transmitted beam is collected by Lens C and imaged onto Camera B, while a polarizing beam splitter (PBS), combined with a half-wave plate (HWP) and a quarter-wave plate (QWP), directs the reflected beam onto Camera A. The data acquisition and processing unit analyzes the transmitted- and reflected-light images to predict the two-axis angular offsets of Mirrors A and B, alongside the axial displacement of Lens C. The electronic control module actuates four motorized mirror mounts (Mirrors A–D; C and D are used in the 9-DOF configuration) and one motorized translation stage (Lens B), as shown in the middle and right panels. The left panel shows a general alignment target, which can be aligned using this cavity-assisted method.}
    \label{fig1}
 \end{figure*}
 
\section{Results}\label{sec2}
\par\medskip
\noindent\textbf{Experimental setup}\par\smallskip
We construct an end-to-end automated alignment system based on a customized ResNet-18 neural network model~\cite{he2016deep}. The schematic of the system is illustrated in Fig.~\ref{fig1}. The setup employs a 150~$\mu$W fiber laser beam ($\lambda$ = 626~nm) coupled into an optical resonant cavity (finesse $\sim$20000) via a mode-matching lens. This lens is mounted on a one-dimensional motorized translation stage. Beam steering is controlled by two tilt mirrors, each mounted on a dual-axis motorized mount. During automated coupling, Camera B (a high-speed industrial camera) records the transmitted spatial intensity profile in video mode while laser frequency is scanned across the fundamental and higher-order cavity resonances. The acquired video frames are then preprocessed into a standardized average intensity projection (AIP) image, which serves as the input for the neural network inference. Simultaneously, Camera A tracks the centroid of the reflected beam, providing auxiliary closed-loop feedback for slow drift compensation. In this implementation, ResNet-18 is used as a lightweight regression engine for the cavity alignment problem. Its residual shortcut connections provide stable gradient propagation and robust feature extraction, while its compact parameter size and fast inference speed satisfy the latency requirement of real-time closed-loop control. The original classification head is replaced by a regression head that directly maps the standardized transmission image to the 5-DOF control state relative to the optimal-coupling reference. Following a single forward pass, the model outputs a predicted 5-DOF control state required for optimal coupling, consisting of four mirror tilt angles and one lens axial displacement. This control vector is transmitted to the actuator controller, which updates the motorized mirror mounts and translation stage. The updated cavity state is then recorded by Camera B, processed and fed back into the network, forming an image--inference--actuation closed loop with a 10~ms end-to-end latency.
\par\medskip
\noindent\textbf{Data-driven structural diagnostics}\par\smallskip

Traditional approaches to structural characterization rely on linearized analytical models (e.g., ABCD ray transfer matrices~\cite{siegman1986lasers}), which assume small deviations from a nominal operating point and break down when navigating the highly nonlinear, multi-parameter response surface of high-finesse cavities. Here, we present a data-driven alternative that trains a neural network to map transmission images to 5-DOF state vectors and subsequently applies PLS~\cite{wold2001pls,rosipal2005overview,mehmood2020comparison} to extract the latent variable that is most closely related to the optical performance, and quantify the sensitivity contribution of each physical degree of freedom.

\begin{figure}[htbp]
    \centering
    \includegraphics[width=\linewidth]{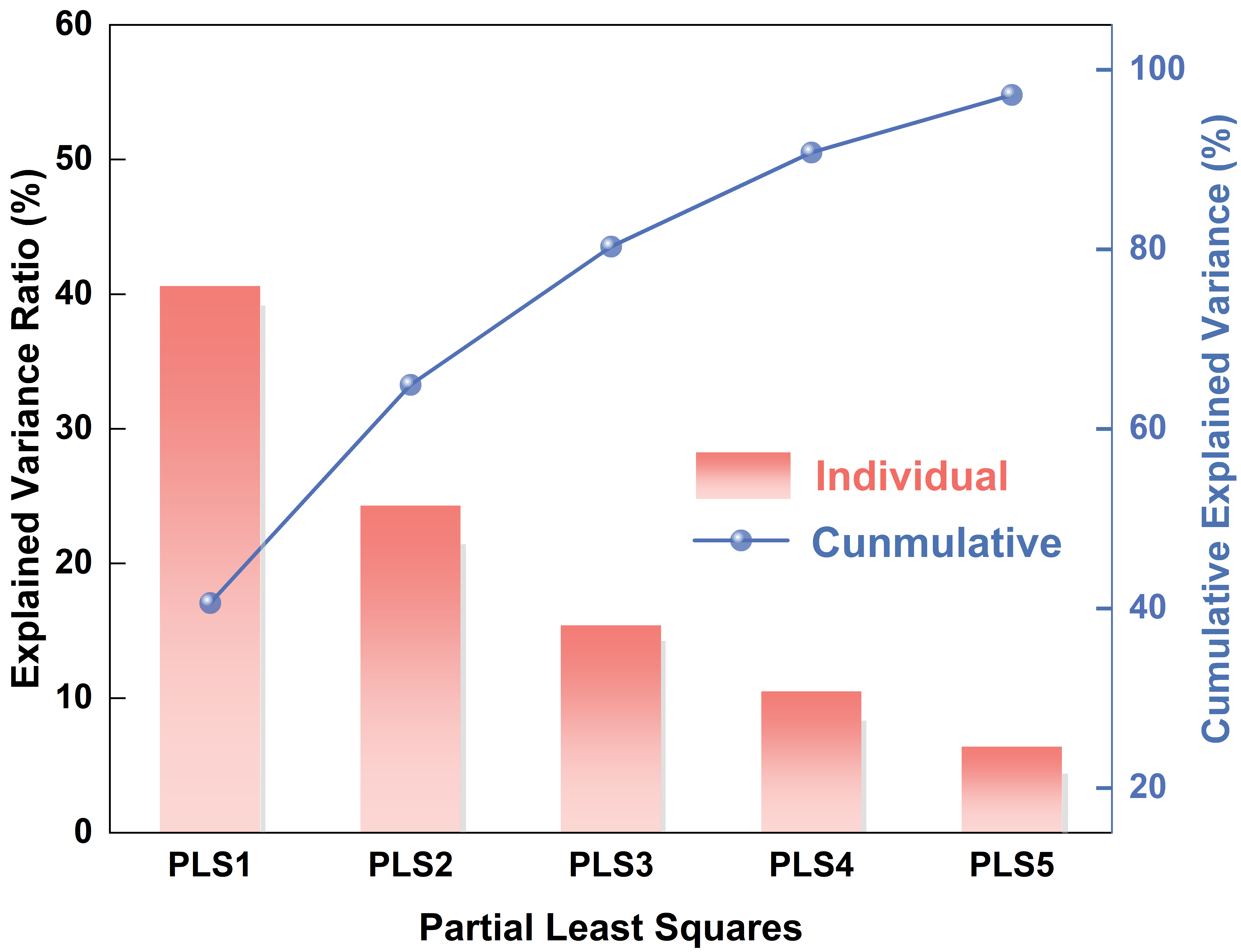}
    \caption{$\mathbf{\vert}$\textbf{PLS variance decomposition of the 5-DOF optical system.} The explained variance ratio spectrum reveals that PLS1 alone accounts for 40.6\% of the purity-relevant control variance, while the top three components (PLS1--PLS3) cumulatively capture 80.3\%. The remaining two PLS directions contribute only 16.9\%, indicating weakly sensitive control directions with comparatively small influence on mode purity.}
    \label{fig2}
\end{figure}

\begin{table}[htbp]
\centering
\caption{Coefficient mapping of the physical degrees of freedom onto PLS1 and PLS2}
\begin{tabular}{lccccc}
\toprule
 & $\boldsymbol{\theta_{x1}}$ & $\boldsymbol{\theta_{y1}}$ & $\boldsymbol{\theta_{x2}}$ & $\boldsymbol{\theta_{y2}}$ & $\boldsymbol{Z\_lens}$ \\
\midrule
\textbf{PLS1} & $0.188$ & $0.546$ & $0.576$ & $0.490$ & $0.307$ \\
\textbf{PLS2} & $-0.511$ & $-0.438$ & $-0.056$ & $0.342$ & $0.653$ \\
\bottomrule
\end{tabular}
\label{tab1}
\end{table}

More critically, the PLS decomposition orders its latent directions by covariance with the fundamental-mode purity, so the first component, PLS1, is by construction the direction most strongly coupled to the optical output.
Within this dominant direction, the eigenvector loadings (Table~1) show that the dominant direction is carried primarily by the downstream mirror angles
($\theta_{x2}=0.576$, $\theta_{y2}=0.490$), together with a substantial contribution from the upstream yaw ($\theta_{y1}=0.546$) and the lens axial
displacement ($Z_{\mathrm{lens}}=0.307$), whereas the upstream pitch ($\theta_{x1}$) contributes the least (0.188). This pattern indicates that
longitudinal focusing and downstream angular steering are combined into a single, coupled control direction in PLS1, rather than acting as independent
adjustments. Notably, $Z_{\mathrm{lens}}$ also loads strongly on the secondary component (0.653 on PLS2), so the lens axial degree bridges both response-driven axes, supporting a structural coupling. Decoupling the lens position from the corresponding angular compensation of the downstream mirrors degrades the fundamental mode purity in routine manual alignment. Consequently, PLS1, as the
covariance-maximizing and structurally coupled direction, provides the primary control pathway for maintaining high fundamental-mode purity and stable cavity coupling.

These results define a data-driven sensitivity hierarchy that directly guides the subsequent active stabilization strategies for the hardware. The highest-priority axis (PLS1, accounting for 40.6\% of the variance) requires joint adjustment of the downstream mirror angles and the lens axial position during coarse alignment. The secondary axis (PLS2, 24.3\% variance) is itself a coupled direction dominated by the lens axial displacement (0.653) together with the upstream mirror angles ($\theta_{x1}$ = $-$0.511, $\theta_{y1}$ = $-$0.438), decoupled from PLS1 to first order, and is therefore suitable for sequential tuning once the PLS1-dominated coarse coupling is established. Together, PLS1 and PLS2 already account for 64.9\% of the explained variance, and the first three components reach 80.3\%; the residual lower-order axes (e.g., PLS4--PLS5, collectively 16.9\%) carry only minor contributions to global alignment. By replacing an unstructured five-dimensional search with this hierarchy-driven protocol, the diagnostic framework enables targeted intervention, a capability that will become increasingly essential as optical platforms scale to higher degrees of freedom.


\begin{figure*}[htbp]
    \centering
    \includegraphics[width=\linewidth]{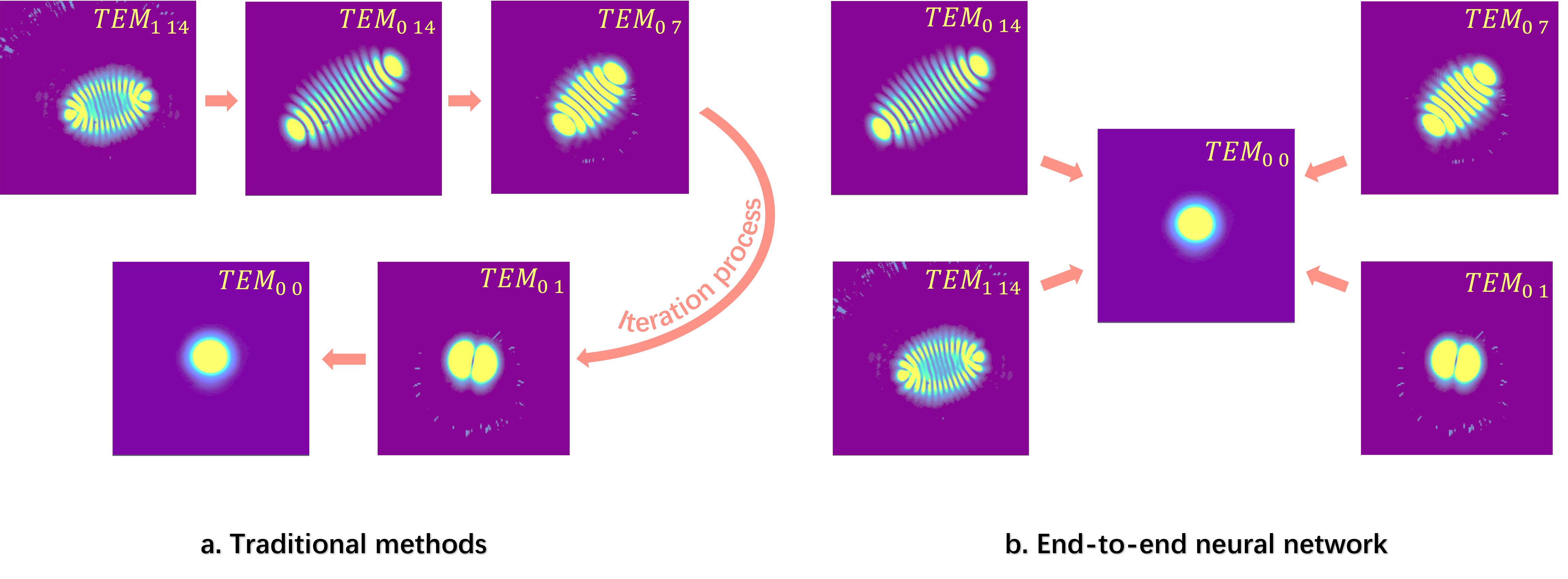}
    \caption{$\mathbf{\vert}$\textbf{ Comparison of alignment paradigms.} \textbf{a.} Iterative trajectories of conventional heuristic search methods (e.g., hill climbing and genetic algorithms) during cavity coupling. The beam profile evolves from higher-order transverse modes toward the fundamental mode (TEM$_{00}$) over multiple iterations. The trajectory is susceptible to local minima and its convergence is sensitive to the initial condition. \textbf{b.} End-to-end neural-network alignment. Provided the input image lies within the training domain, a single forward pass directly predicts a near-optimal 5-DOF control vector for optimal coupling. In many cases, this single inference step moves the system into the high-purity capture basin without intermediate iterations, substantially improving alignment efficiency and robustness. }
    \label{fig4}
\end{figure*}

\par\medskip
\noindent\textbf{Diagnostics-informed autonomous alignment}\par\smallskip
Traditional optimization algorithms often fail to converge when navigating multi-modal parameter landscapes or operating under mechanical noise. For instance, the hill-climbing method exhibits a high convergence failure rate when initial lateral offsets exceed 2~mm (Fig.~\ref{fig4}a). However, the proposed model learns the complex input-output mapping during training, enabling the direct regression of a near-optimal control state via a single forward pass (Fig.~\ref{fig4}b).

The findings outlined above directly inform a diagnosis-guided, high-efficiency hybrid alignment strategy. Specifically, the effective control dimensionality is substantially reduced from the nominal 5-DOF, with PLS1 strongly coupling three key hardware parameters. We therefore deploy the neural network as a global navigator that operates primarily along this principal subspace, bypassing unstructured exploration of the full 5-D space. This global navigation rapidly steers the system into a well-defined capture basin, setting the stage for the subsequent local fine-tuning phase.

\begin{figure}[htbp]
    \centering
    \includegraphics[width=0.9\linewidth]{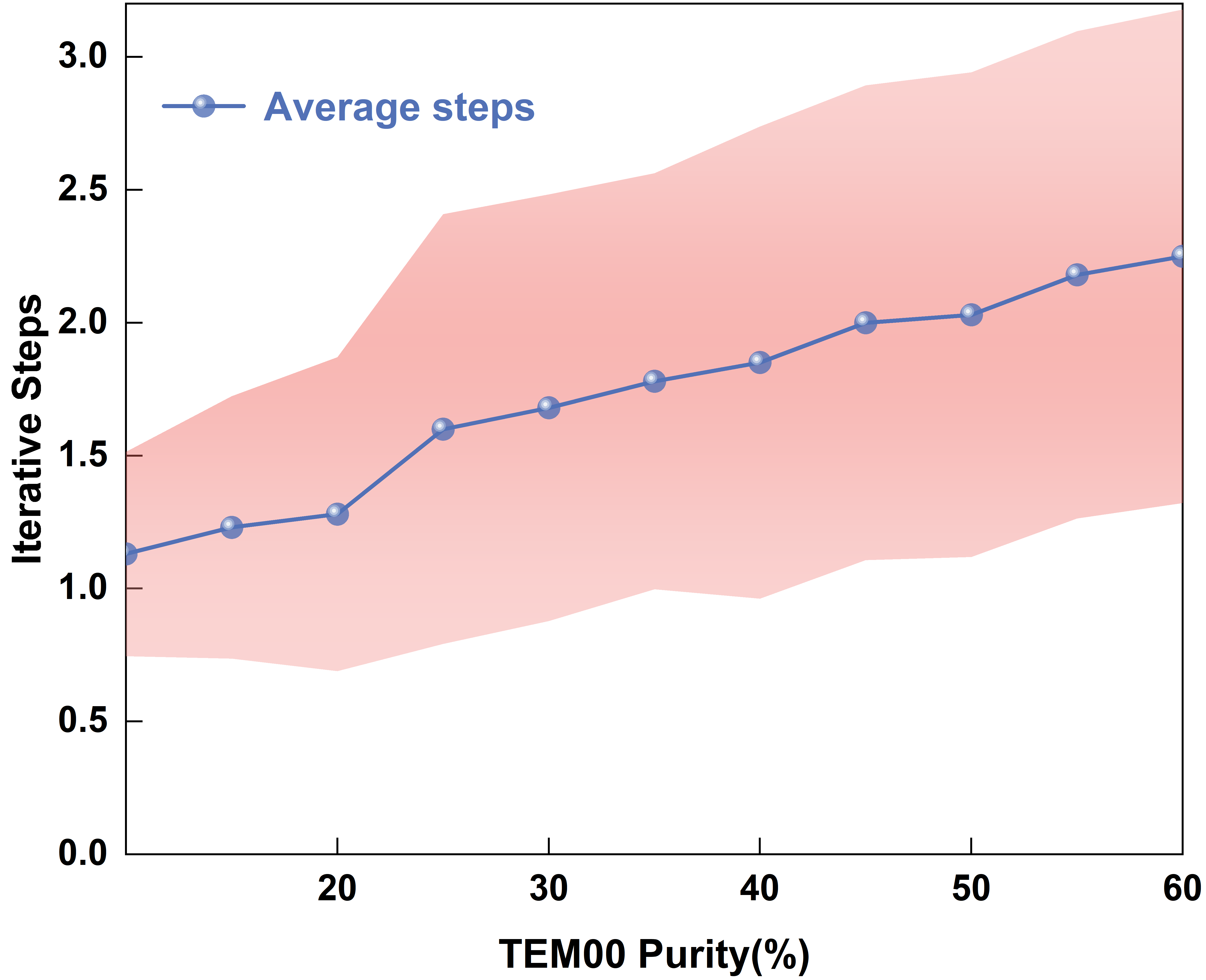}
    \caption{$\mathbf{\vert}$\textbf{ Cold-start recovery performance.} Mean number of iterative steps required to reach a given fundamental-mode (TEM00) purity, averaged over 40 independent trials initiated from randomized gross misalignments. The shaded band indicates the trial-to-trial spread ($\pm$1 standard deviation). The mean trajectory increases monotonically from one iteration at the low purity to $\sim$2.25 iterations at the 60\% purity used for the global-to-local switching threshold, indicating that the global navigation stage reaches the capture basin within a few iterations.}
    \label{fig5}
\end{figure}

The physical closed-loop experiments were conducted under conditions of unavoidable sensor noise and mechanical backlash. We performed 40 independent trials initiated from broadly randomized gross misalignments, far exceeding the 2~mm threshold beyond which hill-climbing fails (Fig.~\ref{fig5}). The results demonstrate that the end-to-end architecture elevates the fundamental mode purity to above 60\% within approximately 2.25 global iterations on average. This ``cold-start'' recovery capability which traverses from near-zero coupling to a well-defined resonance within tens of seconds, stems from the model's implicit learning of the control manifold structure.

Specifically, the transition between the global and local regimes is governed by a predefined purity threshold: once the model-calculated fundamental mode purity exceeds 60\%, the system automatically disengages the neural network's global inference loop and enters a sequential single-axis linear scan.  This handoff is justified because, within the capture basin ($>60$\% fundamental-mode purity), the local response surface becomes sufficiently smooth and weakly coupled that a coordinate-descent scan converges reliably. In this local fine-tuning phase, the PLS loadings determine the scan order of axes and each of the five control degrees of freedom is scanned independently along a narrow linear range while the remaining four are held fixed. For each axis, the system sweeps through a predefined set of discrete steps, continuously monitors the real-time fundamental mode purity, and halts at the position yielding the maximum value. In practice, a single cycle (one pass over all five axes) is typically sufficient to drive the final mode purity above 90\%~\cite{qin2025automated,anderson1984alignment}, with the entire local stage concluding within 45~s. Across all 40 independent trials, this hybrid strategy achieves convergence in every trial.

This composite ``global data-driven coarse alignment + local linear fine-tuning'' architecture resolves the inherent trade-off in which the neural network provides strong global navigation but is limited in the fine-alignment regime by a residual prediction error (the normalized MAE of the 5-DOF parameter predictions is $\sim$0.07), whereas a local linear scan is precise yet globally fragile. The diagnostic hierarchy provides the theoretical justification for this handoff, because within the capture basin ($>60\%$ purity), the response surface becomes sufficiently smooth and weakly coupled that a coordinate-descent scan, reliably converges to the high-purity operating point within the basin.

It is worth noting that, although the neural network possesses the theoretical capacity for ``single-shot'' global coupling and individual experimental trials did achieve optimal alignment in a single forward pass, the statistical average required 2.25 iterations. This discrepancy arises from three physical and systemic constraints. Unavoidable mechanical backlash in the motorized mirror mounts results in execution deviations between the network's predicted coordinates and the actual physical displacement. In states of severe misalignment, the transmitted signal becomes extremely weak and is dominated by Poisson noise, which compromises the model's prediction fidelity. Finally, the deep model inherently carries a residual validation error ($\sim$0.07 normalized MAE), limiting its absolute precision in the fine-alignment regime. A quantitative analysis of this loss evolution is provided below.

Consequently, the neural network model serves as an effective global navigator, rapidly steering the system out of rugged, non-convex regions into a well-defined basin of attraction ($\sim$60\% fundamental-mode purity) over 2 to 3 coarse iterations, after which the diagnostics-informed linear scan ensures deterministic convergence.
\par\medskip
\noindent\textbf{Model evaluation and experimental validation}\par\smallskip

\begin{figure}[htbp]
    \centering
    \includegraphics[width=0.9\linewidth]{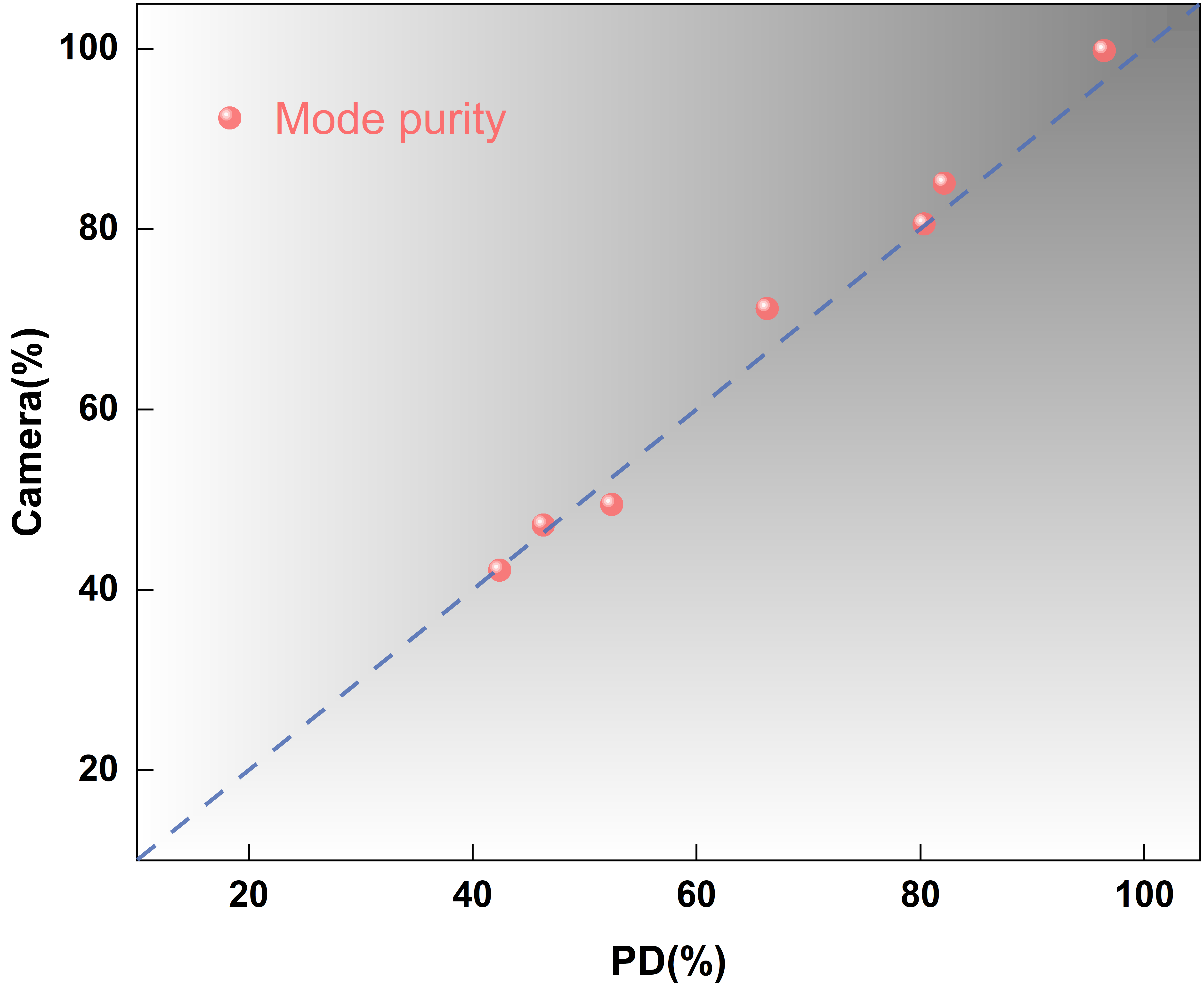}
\caption{$\mathbf{\vert}$\textbf{ Vision model versus electro-optical detection.} Fundamental-mode (TEM00) purity from the proposed vision model (camera B) compared with the photodetector (PD, solid) reference
on the same beam samples. The vision-based purity estimates agree closely with the PD measurements, yielding a mean absolute percentage error of 3.81\%.}
    \label{fig6}
\end{figure}

In this framework, the customized ResNet-18 model demonstrated high prediction fidelity and generalization robustness for the 5-DOF parameter regression task. To evaluate generalization performance, the offline dataset was randomly partitioned into training (85\%) and validation (15\%) subsets. The model achieved a low validation MAE of $\sim$0.07 for the 5-DOF predictions, with both training and validation losses decreasing monotonically and showing no divergence, indicating no overfitting (see Supplementary Note  for training
dynamics). For ground-truth mode purity, we employed a standard photodetector (PD) coupled with oscilloscope temporal analysis of transient resonance peaks (Supplementary Note ). As illustrated in Fig.~\ref{fig6}, the vision-based fundamental-mode purity estimates agree closely with the PD measurements, yielding a mean absolute percentage error of 3.81\%, within the acceptable experimental tolerance ($\pm$5\%).

To further assess alignment efficiency under cold-start conditions, we benchmarked the proposed protocol against standard Genetic Algorithms (GA)~\cite{vorontsov1998stochastic} and stochastic parallel gradient descent (SPGD)~\cite{hu2020adaptive}. Under equivalent initial misalignment, conventional heuristic methods typically require hundreds of iterations and are prone to stochastic trapping in local optima, whereas the proposed protocol, by leveraging the learned latent feature structure, reaches the target purity in a few iterations,
a convergence-speed improvement of more than two orders of magnitude (Supplementary Note for trajectory comparisons and hyperparameters). These results demonstrate the potential of data-driven frameworks to efficiently navigate multi-DOF optical alignment tasks that remain challenging for traditional iterative methods.
\par\medskip
\noindent\textbf{Dimension scalability and training data requirements}\par\smallskip
To assess the scalability of the proposed framework across higher-dimensional optical platforms, we systematically tested three representative DOF configurations. Beyond the baseline 5-DOF system, we implemented a reduced 2-DOF configuration (bi-axial tilt of a single mirror) as a lower-bound reference, and an extended 9-DOF configuration by inserting an additional pair of upstream steering mirrors (Mirrors C and D in Fig.~\ref{fig1}), thereby introducing four angular control parameters ($\theta_{x3}$, $\theta_{y3}$, $\theta_{x4}$, $\theta_{y4}$). For the 9-DOF configuration, automated data acquisition was performed via coordinated motorized sweeps across all nine axes using a quasi-random sampling strategy, requiring approximately 20 hours of continuous operation to generate a sufficiently dense training dataset ($N \approx 9000$ samples).

\begin{figure}[htbp]
    \centering
    \includegraphics[width=0.9\linewidth]{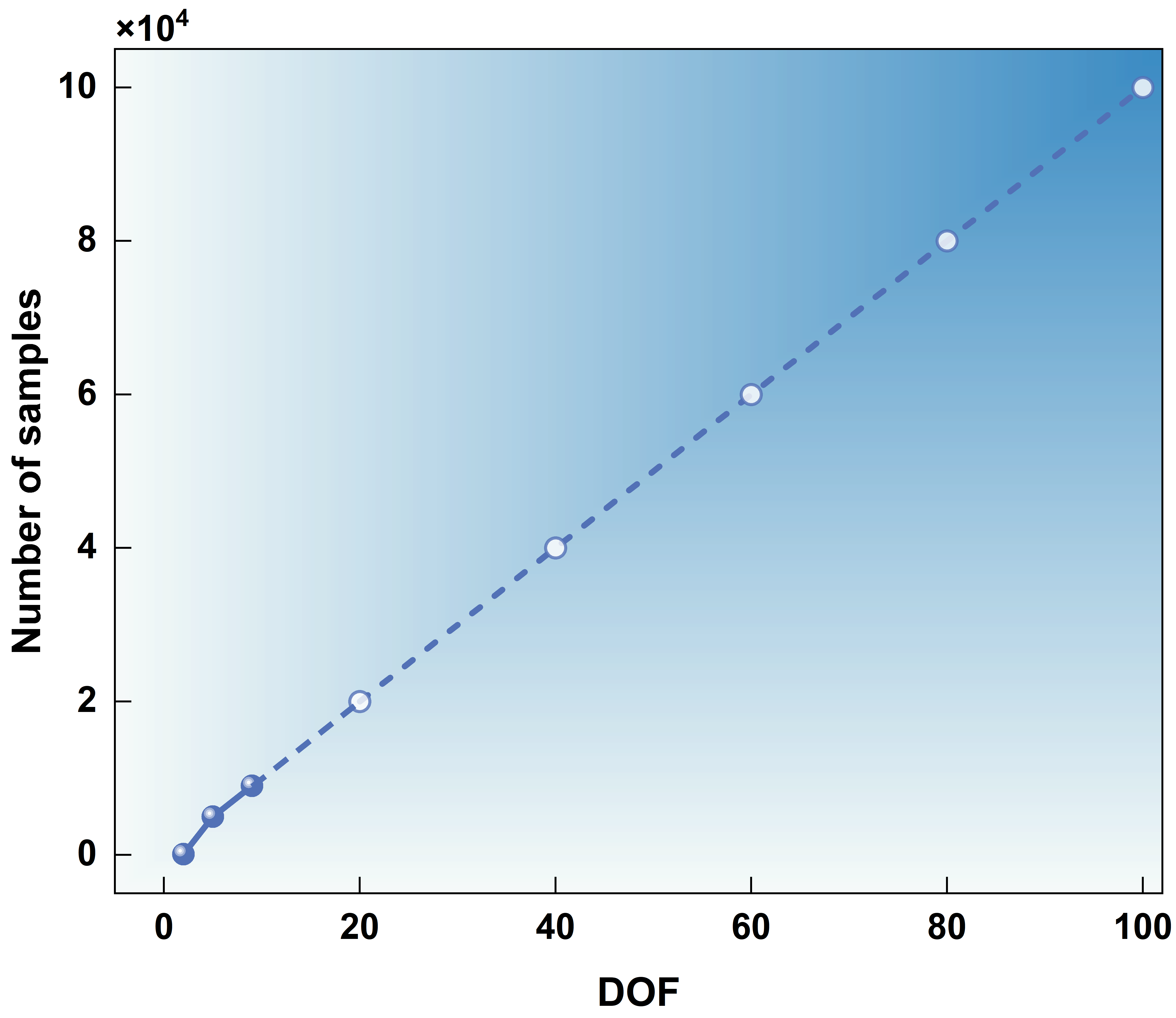}
    \caption{$\mathbf{\vert}$\textbf{ Training data requirements versus system degrees of freedom (DOF).} The first three data points  (2-DOF, 5-DOF, and 9-DOF) are experimentally validated configurations, all achieving validation loss below 0.1. The remaining data are theoretical extrapolations based on the linear scaling law $N\approx 1000k$, derived from architecture-aware analysis of the pretrained ResNet-18 regression head. The approximately linear growth contrasts with the exponential scaling of conventional grid-search approaches, suggesting practical feasibility of extending the framework to platforms with up to 100~DOF.}
    \label{fig7}
\end{figure}

As shown in Fig.~\ref{fig7}, the end-to-end ResNet-18 model maintains a validation MAE consistently below the 0.1 threshold across all three configurations: 0.008 for 2-DOF (trained on $\sim$100 samples), 0.063 for 5-DOF ($\sim$5000 samples), and 0.042 for 9-DOF ($\sim$9000 samples). The 2-DOF case attains the lowest error with nearly two
orders of magnitude fewer samples than the higher-DOF cases, as its compact two-dimensional parameter space is fully spanned by a minimal dataset and the regression task is accordingly close to its error floor. The 9-DOF error (0.042) falling below the 5-DOF value (0.063), rather than rising monotonically with dimensionality, reflects the larger 9-DOF training set together with the comparatively weak coupling of the added upstream mirrors to
the cavity output. For the 5-DOF and 9-DOF configurations, the required training set size follows an approximately linear scaling, $N \approx 1000k$ (where $k$ denotes the DOF count), which maintains sufficient coverage of the parameter space despite its combinatorial growth. This indicates that the deep inversion architecture retains robust predictive fidelity as system complexity increases. Crucially, for these higher-DOF configurations the training data requirement grows approximately linearly with $k$, in contrast to the exponential sampling complexity inherent to conventional grid-search methods. The theoretical basis for this linear scaling rests on the dimension-independent task-specific capacity of the frozen pretrained backbone together with a quasi-random sampling strategy at fixed per-axis resolution, and is detailed in the Methods.

\section{Discussion}\label{sec4}
ResNet-18 was adopted as the core inverse engine, prioritizing global state recovery across the control landscape over sub-micron fine-tuning. Residual connections ensure stable gradient flow under low-SNR conditions, enabling robust recovery from severe cold-start distortions. This enables the network to steer the system into the capture basin, complementing the diagnostics-informed local scan, which is precise yet globally fragile. Furthermore, this architecture meets the low-latency requirements of hardware closed-loop control with a 10~ms end-to-end latency.

Scaling this framework from single-cavity alignment to large-scale quantum optical arrays requires automating the entire control pipeline. To enhance modular scalability and human-machine interaction, integrating standardized control interfaces (e.g., the Model Context Protocol, MCP) with LLM-based orchestration agents is a promising direction toward autonomous optical laboratories, in which such an agent would invoke the inverse model and the PLS diagnostics as standardized tools. Optical physicists could then issue high-level
natural-language commands, and the agent would autonomously schedule data acquisition, trigger adaptive probing, manage model retraining, and deploy updated inference policies in real time, moving toward fully goal-driven fine optical experiments.

The proposed ``End-to-End Inversion + Coupling Dimensionality Reduction'' diagnostic paradigm provides a data-driven diagnostic framework for complex optical hardware with continuous parameter spaces. For future macroscopic phase-controlled platforms spanning hundreds of distinct physical nodes, such as massively scaled cold-atom qubit arrays or orbital angular momentum communication meshes, the rapid geometric expansion of internal structural coupling renders traditional deterministic modeling inadequate. Our framework delivers an efficient, robust closed-loop alignment protocol alongside a systematic diagnostic suite for sensitivity analysis. By analyzing the extracted the main sensitive direction and the corresponding eigenvalues, dominant structural couplings can be inversely resolved from high-dimensional mechanical parameter spaces. This framework offers a path away from iterative ``blind empiricism'' in massive optical tuning, establishing a scalable, data-driven methodology for addressing the engineering complexity of next-generation free-space quantum optical systems.

\section*{Methods}\label{sec3}
\par\medskip
\noindent\textbf{Ill-posedness of the forward physical model and inverse problem}\par\smallskip
The optical resonant cavity is controlled via five degrees of freedom: the tilt angles of two steering mirrors ($\theta_{x1}, \theta_{y1}, \theta_{x2}, \theta_{y2}$) and the axial position of the mode-matching lens ($Z_{\mathrm{lens}}$), collectively defining the state vector $\boldsymbol{X} \in \mathbb{R}^5$. The system observable is the two-dimensional spatial intensity profile $I(\mathbf{r})$ recorded by a camera at the transmission port. The forward mapping $\boldsymbol{X} \to I(\mathbf{r})$ arises from the nonlinear relationship between alignment parameters and transmitted light intensity profile (and hence mode purity), governed by cavity eigenmode interference and detection noise, rendering it a highly non-injective function.

Automated alignment constitutes an inverse problem: recovering a near-optimal control vector $\boldsymbol{X}^*$ from a single-shot intensity measurement $I(\mathbf{r})$. However, because intensity imaging discards phase information and the five control parameters exhibit strong parametric coupling, this inverse problem is severely ill-posed. Conventional optimization methods (e.g., gradient descent or heuristic searches) rely on iterative trial-and-error in this non-convex, rugged parameter landscape. They are prone to stagnation at local optima, and the number of iterations required for convergence grows exponentially with dimensionality. To address these challenges, we employ a deep neural network to learn a direct, end-to-end mapping $I(\mathbf{r}) \rightarrow \boldsymbol{X}$, thereby bypassing iterative search procedures and mitigating convergence failures in cold-start scenarios.
\par\medskip
\noindent\textbf{Manifold structure and embedding in optical phase space}\par\smallskip
As discussed above, the transmitted intensity $I(\mathbf{r})$ is determined by the 5-DOF control vector $\boldsymbol{X}$. Although raw sensor images reside in a high-dimensional pixel space (e.g., $\mathbb{R}^{m \times n}$), the physical constraints imposed by Gaussian beam propagation and cavity eigenmode orthogonality restrict all physically valid optical responses to a low-dimensional, highly structured nonlinear manifold $\mathcal{M}$~\cite{fefferman2016testing}.
\begin{equation}
    \mathcal{M} = \{ I(\mathbf{r}) = \mathbf{g}(\boldsymbol{X}) \mid \boldsymbol{X} \in \Omega \} \subset \mathbb{R}^{m \times n}
\end{equation}
Here, $\mathbf{g}$ denotes the forward physical mapping operator governed by cavity optics, and $\Omega \subset \mathbb{R}^5$ represents the bounded parameter domain defined by the mechanical limits of the actuators. The intrinsic dimensionality of $\mathcal{M}$ is substantially lower than the ambient pixel dimension, providing the theoretical foundation for learning-based inversion.

We employ a convolutional neural network (CNN) to learn an approximate inverse mapping $\mathbf{g}^{-1}$ that projects high-dimensional image observations directly into the 5-DOF control parameter space. Geometrically, training the network approximates a smooth mapping that unfolds the low-dimensional optical manifold into a more linear latent representation. This facilitates the implicit disentanglement of underlying physical parameters, as consistent with the PLS-based sensitivity analysis in the Results.
\par\medskip
\noindent\textbf{Physical feature separation mechanism}\par\smallskip
In misalignment regimes where multiple parameters are strongly coupled, the neural network implicitly decouples each control degree of freedom. This capability emerges from the network's multi-scale receptive fields, which effectively capture asymmetrical intensity modulations, higher-order transverse mode evolutions, and envelope gradient variations induced by physical misalignment.

Angular misalignment (tilt) breaks the transverse spatial symmetry of the intra-cavity field, exciting odd-order higher transverse modes (e.g., $HG_{10}/HG_{01}$). Coherent interference among these asymmetric modes generates complex interference patterns on the transmission plane, manifesting as energy redistribution and distinct center-of-mass spatial walk-offs~\cite{saleh2019fundamentals}. The walk-off direction directly encodes the specific pitch/yaw deviation of the hardware. CNN architectures naturally learn to detect such asymmetrical gradients, with shallow convolutional layers acting as implicit edge detectors that respond strongly to localized asymmetrical spatial derivatives~\cite{zeiler2014visualizing}. Through hierarchical aggregation of localized receptive fields, the network establishes a robust mapping between first-order positional asymmetries and their corresponding angular tilt parameters in the latent space.

In contrast to symmetry-breaking angular deviations, axial displacement of the focusing lens induces a longitudinal shift of the fundamental Gaussian beam waist. According to paraxial propagation theory, the beam waist radius evolves as~\cite{siegman1990new}
\begin{equation}
   \omega(z) = \omega_0 \sqrt{1 + \left( \frac{z - z_0}{z_R} \right)^2},
\end{equation}
where $\omega_0$ is the minimum beam waist radius, $z_R$ is the Rayleigh length, and $z_0$ denotes the waist location. The corresponding intensity distribution on the observation plane is $I(r, z) \propto \exp\left(-2r^2/\omega^2(z)\right)$. Since this paraxial perturbation preserves rotational symmetry (no odd-order mode coupling), the optical field evolution manifests solely as isotropic radial scaling and a ``breathing effect'' of the global envelope. The network captures this symmetric modulation through deep layers with large receptive fields and global pooling, associating macroscopic variance and second-order spatial moment statistics with the lens axial position.

Leveraging the collaborative extraction of multi-scale features, the network constructs an approximately factorized physical representation in the latent space. Let $\mathbf{z}_\theta$ and $\mathbf{z}_Z$ denote the angular and axial latent feature vectors, respectively. By minimizing the regression loss under the constraint of a compact regression bottleneck architecture, the network implicitly reduces the statistical dependence between $\mathbf{z}_\theta$ and $\mathbf{z}_Z$, yielding near-orthogonal angular and axial subspaces~\cite{chen2018isolating,tishby2015deep}:
\begin{equation}
        \mathbf{z}_\theta^T \cdot \mathbf{z}_Z \approx 0
\end{equation}

It must be emphasized that this multi-DOF decoupling mechanism constitutes a data-driven ``soft disentanglement''. Rather than imposing hard algebraic independence priors, the model naturally converges to an efficient statistical coding scheme that separates coupled physical variances through massive training data.
\par\medskip
\noindent\textbf{Partial least squares analysis}\par\smallskip
To quantify the parameter coupling structure and link it to system optical performance, we apply PLS to the ensemble of network-predicted 5-DOF control vectors, jointly with the corresponding output responses (i.e., fundamental-mode purity). The first PLS latent variable corresponds to the direction that maximizes the covariance between the control space and the response space, representing the dominant sensitivity axis that most strongly co-varies with the optical response. Subsequent latent variables (PLS2, PLS3, …) are sequentially extracted from the residual subspaces of both the control vectors and the responses, capturing secondary covariance directions in a descending order of explanatory power for the output. This yields a response-oriented low-dimensional representation of the coupled 5-DOF control structure. Specifically, we perform Z-score normalization on the control vectors and the response variable, and then execute the iterative PLS algorithm to extract latent variables sorted by their covariance contributions. By retaining only the dominant components (e.g., PLS1–PLS3), we isolate the primary regulating modes that govern the system's macroscopic alignment behavior and determine the final optical performance, effectively reducing the control dimensionality while preserving the core response-sensitive structural characteristics of the misalignment data (see Supplementary Note for rigorous mathematical derivations).
\par\medskip
\noindent\textbf{Data acquisition and processing}\par\smallskip
In this experimental closed-loop, a customized ResNet-18 architecture was deployed. To ensure robust generalization across the nonlinear manifold linking image features to actuator positions, the system was first manually aligned to a near-optimal fundamental mode resonance. Subsequently, algorithmic random perturbations were injected across the five control degrees of freedom, driving the hardware into highly coupled misalignment regimes.Camera B synchronously recorded the transmitted intensity sequences with Camera A providing visual feedback for precise positional callback. Once acquisition was completed, the actuators were returned to the initial reference positions. This protocol yielded a training dataset of 5,000 labeled image-parameter pairs, spanning a broad range of multidimensional coupling configurations. The dataset was subsequently randomly partitioned into training (85\%) and validation (15\%) subsets.

Due to the transient nature of the swept signals and low photon counts under severe misalignment, raw frames were pre-processed to mitigate Poisson noise. Through temporal frame synchronization, 2D spatial integration, and intensity normalization (detailed in Supplementary Note), standardized AIP maps were generated as model inputs for both offline training and online inference.

Furthermore, to obtain reference labels for fundamental mode purity, an independent image-analysis pipeline was executed in parallel. This pipeline employed classical spatial metrics (e.g., circularity index, second-order moments) and temporal trajectory tracking to estimate the TEM\textsubscript{00} fraction. The computational framework for this purity estimation is provided in Supplementary Note. These image-derived labels were validated against photodetector (PD) benchmarks, achieving a MAPE of 3.81\% (see Fig.~\ref{fig6}), confirming their reliability for supervised training. 
\par\medskip
\noindent\textbf{Theoretical analysis of dimension scalability}\par\smallskip
The approximately linear scaling of training data requirements observed empirically ($N \approx 1000k$) can be grounded dimension-independent capacity and the per-axis sampling strategy employed in this work. The framework adopts a pretrained ResNet-18 as its convolutional backbone. Pretrained on large-scale natural image datasets, this backbone learns generic visual primitives such as edges, shapes, and intensity textures, which generalize effectively to optical beam pattern analysis. Functionally, the backbone compresses high-dimensional pixel inputs into a compact 512-dimensional feature vector, acting as a fixed implicit dimensionality reducer. Crucially, the backbone parameters ($\sim$11.2 million) remain frozen during task-specific fine-tuning and thus do not contribute to the sample complexity of the alignment problem.

The only trainable component is the terminal regression head. Its parameter count, $P_{\text{head}}$, decomposes into two structurally distinct terms: a fixed cost and a variable cost. The fixed cost corresponds to the first fully connected layer ($512 \rightarrow 256$), which transforms image features into physical mapping features, yielding $512 \times 256 + 256 = 131{,}328$ parameters. This term is architecture-determined and independent of the control dimensionality $k$. The variable cost arises from the second fully connected layer ($256 \rightarrow k$), which maps the learned physical features onto the $k$-dimensional control vector, contributing $256 \times k + k = 257k$ parameters.
Thus, the total trainable parameter count is
\begin{equation}
P_{\text{head}} = 131{,}328 + 257k.
\end{equation}
At $k = 100$, the variable term amounts to only $25700$,  representing less than 20\% of the fixed term and indicating that the incremental parameter cost remains negligible relative to the backbone's fixed capacity. Consequently, the network's effective learning capacity does not scale combinatorially with $k$, mitigating the overfitting risk that typically plagues high-DOF regression tasks.

Since the model capacity is effectively dimension-independent, the sample complexity is primarily governed by the requirement to adequately cover the $k$-dimensional parameter manifold. Because each of the 
k axes is independently swept with a fixed per-axis point budget, the total sample count grows linearly with $k$
\begin{equation}
N_{\text{min}} \propto k.
\end{equation}
This empirical trend is validated by the 5-DOF ($N \approx 5.0 \times 10^3$) and 9-DOF ($N \approx 9.0 \times 10^3$) configurations. Extrapolating from this linear scaling relation, platforms with 50--100 DOF would require datasets on the order of $5 \times 10^4$ to $10^5$ samples. With parallelized multi-axis actuation, the per-sample cost is kept near-constant, so the total acquisition time scales primarily with the sample count N. Modern motorized stages with parallelized control and sub-second settling can render this feasible within practical automated campaigns. This supports the engineering viability of extending the framework to next-generation high-DOF quantum optical platforms, provided the underlying control manifold remains low-dimensional due to physical constraints.



\section*{Acknowledgments}
This research was supported by the Guangdong-Hong Kong-Macau Greater Bay Area Quantum Science Center and funded through the Guangdong Province Quantum Science Strategic Initiative (Grant No.~GDZX2302005), Guangzhou-HKUST(GZ) Joint Funding Program (No.~2025A03J3783) and Guangzhou Municipal Science and Technology Project (No.~2025A04J7077, No.~2025A03J3861). J.~F.~C. acknowledges the National Natural Science Foundation of China (NSFC) through Grants No.~92476102, and Guangdong Key Project under Grant No. 2022B1515020096.

\section*{Author Contributions}
P.C. proposed and supervised this project. P.Z. carried out the entire
experiment, including establishing the experimental system as well as collecting and analyzing the data. P.C. and J.F.C. provided guidance for the research work. All authors contributed to the discussion, analysis, and interpretation of the results, and co-wrote the manuscript.

\section*{Competing Interests}
The authors declare no competing interests.

\section*{Data availability}
All data necessary to support the conclusions of the paper are available upon reasonable request.

\bibliography{reference}

\end{document}